\documentclass[11pt]{article}
\usepackage[margin=1in]{geometry}
\usepackage{parskip}
\usepackage{natbib}
\usepackage{booktabs}
\usepackage{multirow}
\usepackage{array}
\usepackage{longtable}
\usepackage{tabularx}
\usepackage{makecell}
\usepackage{float}
\usepackage{tikz}
\usepackage{hyperref}
\usepackage{graphicx}
\usepackage{amsmath}
\usepackage{amssymb}
\usepackage{caption}
\usetikzlibrary{arrows.meta, positioning, calc}
\begin{document}

\title{Causal Fairness Analysis of ADHD Status and High School STEM Outcomes}
\author{Shuhan (Alice) Ai \\
\textit{School of Education and Information Studies} \\
\textit{University of California Los Angeles, 90095, USA} \\
\href{mailto:aishuhan@ucla.edu}{aishuhan@ucla.edu}}
\date{}
\maketitle

\section{Study Motivation and Research Questions}

Science, technology, engineering, and mathematics (STEM) careers represent an important pathway for upward mobility among people with disabilities \citep{Meyer2017}, but the STEM potential of youth with cognitive disabilities is dismissed by the assumption that these students are inherently unable to succeed. This deficit framing is misleading: an Attention-Deficit / Hyperactivity Disorder (ADHD) diagnosis is not based on low IQ \citep{Matson2010}, and the lower achievement observed among these students reflects not individual factors alone, but the joint product of individual differences and social processes, including stigma, reduced access to advanced coursework, and lower expectations.

Beyond achievement, disability labels may also shape how teachers perceive a child's potential \citep{Owens2020} and can lead to curricular marginalization \citep{Shifrer2013}. Moreover, because cognitive disabilities lack objective biomarkers, these labels are disproportionately assigned to youth from lower-SES and racial minority backgrounds \citep{Saatcioglu2019}. A recent study conducted by \citet{Shifrer2021} found that youth with medicated ADHD, who tend to come from more advantaged backgrounds, show the highest STEM achievement among all disability groups, yet still lag behind non-disabled peers in science identity. However, existing studies rely on descriptive or regression-based methods that cannot unravel the causal mechanisms of these disparities. 

This study aims to address this gap by applying the Causal Fairness Analysis (CFA) framework \citep{Plecko2024}, which is grounded in Pearl's Structural Causal Model \citep{Pearl2009, Pearl2021}, to decompose the observed disparity in science identity and high school STEM GPA between students with ADHD and non-ADHD into three components: confounded (spurious), indirect, and direct effects. In doing so, this study provides a causal perspective on how a neurodevelopmental health condition translates into educational inequality. Three research questions guided this study:

\begin{enumerate}
    \item To what extent does ADHD status contribute to disparities in high school STEM outcomes (science identity and STEM GPA), and through what causal pathways (confounding, indirect, direct) does this disparity operate?
    \item How does the total causal effect of ADHD on STEM outcomes vary across student demographic subgroups defined by race/ethnicity, socioeconomic status, and educational expectations?
    \item To what extent does the direct causal effect of ADHD on STEM outcomes, after removing the influence of mediating academic pathways, differ across students' demographic backgrounds?
\end{enumerate}

The first question addresses the total variation decomposition, decomposing the overall ADHD outcome gap into its constituent causal channels. The second question investigates heterogeneity in the conditional average treatment effect (CATE) across demographic subgroups. The third question examines heterogeneity in the counterfactual direct effect (ctf-DE), isolating the portion of the ADHD penalty that bypasses the measured mediators.

\section{Data and Variables}

\subsection{Study Design Overview}

This study employed a six-step analytic pipeline grounded in the CFA framework \citep{Plecko2024} to investigate how ADHD status shapes high school STEM outcomes. The full analysis pipeline is presented in Appendix~A. Drawing from the High School Longitudinal Study of 2009 (HSLS:09), I constructed two outcome-specific datasets, one for science identity and one for STEM GPA, each organized according to the Standard Fairness Model (SFM) with variables assigned to the protected attribute ($X$), confounders ($Z$), mediators ($W$), and outcomes ($Y$). Missing data ranged from 1\% to 20\% across variables, with an average rate less than 5\%. To address missingness, I applied multiple imputation by chained equations (MICE) with 10 imputations.

The analysis proceeded in three phases. First, I provided descriptive context by comparing covariate and mediator distributions between ADHD and non-ADHD students using standardized mean differences (SMD). Second, I addressed the three research questions sequentially: decomposing the total variation in each outcome into confounded, indirect, and direct effects using semiparametric debiased estimation; estimating heterogeneity in the total causal effect across demographic subgroups using generalized random forest; and examining heterogeneity in the direct causal effect using the one-step debiased estimator with cross-fitting. Third, I assessed robustness through confounder sensitivity analysis (\texttt{sensemakr}) and propensity score overlap trimming. Methodological details for each step are described in the corresponding analysis sections below.

\subsection{Data Source and Sample}

This study used the HSLS:09, a nationally representative survey conducted by the National Center for Education Statistics (NCES) that followed approximately 23,000 U.S.\ students from 9th grade through post-secondary education. I drew on three waves: the base year (9th grade, 2009) for baseline covariates, the first follow-up (11th grade, 2012) for mediating variables and the science identity outcome, and the 2013 college update for the STEM GPA outcome. This longitudinal structure ensures that confounders, mediators, and outcomes are temporally ordered in accordance with the causal model.

The analytic samples included all students with non-missing ADHD diagnosis status and the relevant outcome. The science identity sample comprised 13,760 students, and the STEM GPA sample comprised 14,600 students. Across both samples, students with ADHD were disproportionately male and White. These demographic patterns were consistent with national estimates that approximately 12\% of U.S.\ adolescents have a cognitive disability \citep{OSEP2015}.

\subsection{Variables}

\begin{figure}[htbp]
    \centering
    \includegraphics[width=0.45\textwidth]{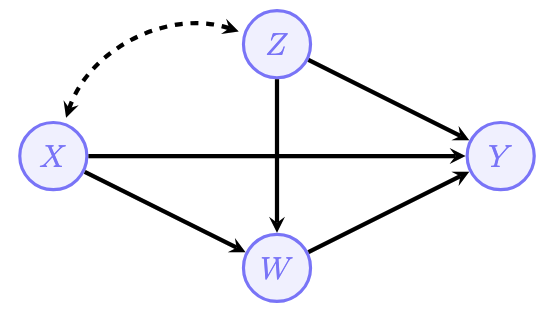}
    \caption{Standard Fairness Model (SFM)}
    \label{fig:sfm}
\end{figure}

As shown in Figure~\ref{fig:sfm}, variables were organized according to the SFM framework \citep{Plecko2024} into four groups: protected attribute ($X$), confounders ($Z$), mediators ($W$), and outcomes ($Y$). Full variable definitions and coding details are provided in Appendix~B.

The protected attribute $X$ was a binary indicator of ADHD diagnosis (1 = diagnosed, 0 = non-ADHD). Confounder block $Z$ captured baseline characteristics not causally influenced by ADHD status: race/ethnicity, sex, family socioeconomic status (SES quintile), 9th-grade science identity, and highest educational expectation. Mediator block $W$ captured 11th-grade STEM learning experiences that ADHD may influence, including students' enrollment in Algebra~II, Geometry, and Pre-calculus, and participation in a science club or study group. The outcome $Y$ was measured in two ways: high school STEM GPA as a performance-based measure, and 11th-grade science identity as an identity-oriented measure.

\section{Analysis}

\subsection{Causal Modeling}

Using the SFM constructed in Figure~\ref{fig:sfm}, I investigated how the observed ADHD-outcome disparity arises through three channels: (1) the direct channel ($X \rightarrow Y$), whereby ADHD status affects STEM outcomes holding all observed confounders and mediators constant; (2) the indirect channel ($X \rightarrow W \rightarrow Y$), capturing the effect transmitted through ADHD's influence on educational engagement; and (3) the spurious channel ($X \leftarrow Z \rightarrow Y$), capturing non-causal variation due to demographic differences between ADHD and non-ADHD students. The first two channels represent causal mechanisms, while the third reflects confounding---group differences that would exist even in the absence of any causal effect of ADHD.

\subsection{Descriptive Balance Diagnostics}

Before estimating the causal effects, I assessed baseline imbalance between ADHD and non-ADHD students. For each confounder in $Z$ and mediator in $W$, I computed SMDs between the two groups---the difference in group means divided by the pooled standard deviation for continuous variables, and SMDs on indicator-coded levels for categorical variables. Following the traditional rule of thumb, I flagged variables with $|\text{SMD}| > 0.10$ as meaningfully imbalanced \citep{ZZhang2019}. These diagnostics served a dual purpose. Substantively, they characterized how ADHD and non-ADHD students differed prior to causal adjustment. Methodologically, they foreshadowed the decomposition: large imbalances in confounders ($Z$) suggested that a substantial portion of the raw outcome gap may be spurious, while imbalances in mediators ($W$) signaled potential indirect pathways.

\subsection{Total Variation Decomposition}

To address the first question, I decomposed the total variation (TV) in each outcome using the \texttt{fairness\_cookbook} implementation of the CFA framework \citep{Plecko2024}. The TV captures the overall observed gap between groups:
\begin{equation}
    \text{TV}_{x_0, x_1}(y) = E\left[Y \mid X = x_1\right] - E\left[Y \mid X = x_0\right]
    \label{eq:tv}
\end{equation}

TV is an associational ($L_1$) quantity under Pearl's Causal Hierarchy (PCH) that does not invoke any intervention. It reflects differences in the composition of units across the two groups, not necessarily any causal effect of $X$ on $Y$. Unlike the total effect $\text{TE}_{x_0, x_1}(y)$, which captures only causal variation flowing from $X$ through pathways (1) and (2), the TV additionally captures back-door (spurious) variation transmitted through confounders $Z$. Under the SFM, the TV decomposes into three components \citep{Plecko2024, Zhang2018}:
\begin{equation}
    \text{TV}_{x_0, x_1}(y) = \underbrace{x\text{-DE}_{x_0, x_1}(y \mid x_0)}_{\text{direct}} - \underbrace{x\text{-IE}_{x_1, x_0}(y \mid x_0)}_{\text{indirect}} - \underbrace{x\text{-SE}_{x_1, x_0}(y)}_{\text{spurious}}
    \label{eq:tvdecomp}
\end{equation}

Each component maps onto one of the three pathways identified above. The $x$-specific direct effect ($x$-DE) captures the residual disparity that would remain if ADHD and non-ADHD students had identical background and educational engagement profiles, the effect transmitted along $X \rightarrow Y$ only. The $x$-specific indirect effect ($x$-IE) captures how much of the gap is transmitted because ADHD affects STEM learning experiences (e.g., advanced coursework, science club participation), which in turn shape STEM outcomes, the $X \rightarrow W \rightarrow Y$ pathway. The $x$-specific spurious effect ($x$-SE) isolates variation through non-causal backdoor paths $X \leftarrow Z \rightarrow Y$: if ADHD diagnosis rates vary by SES, and SES independently affects STEM outcomes, the two groups would exhibit different outcomes even under identical treatment because they are drawn from different confounder strata. Formal counterfactual definitions of each component are provided in Appendix~C.

The decomposition operates at Layer~3 (counterfactual) of the PCH because $x$-DE and $x$-IE involve nested counterfactuals, reasoning about outcomes under one treatment status while holding mediators at the values they would have taken under a different treatment status.

Estimation proceeded via semiparametric debiased (doubly robust) estimation with cross-fitting. This approach combines flexible machine learning models for the nuisance parameters (outcome regression and propensity scores) with bias-correction terms that ensure valid inference even when individual nuisance models are imperfect. Cross-fitting partitions the data into folds, estimating nuisance parameters on held-out folds to avoid overfitting bias. I computed 95\% confidence intervals and reported the proportion of TV attributable to each component.

\subsection{Conditional Average Treatment Effect (CATE) Estimation}

To address the second research question, I estimated heterogeneity in the total causal effect using generalized random forest (GRF; \citealp{Athey2019}). Where the TV decomposition provided a single population-level estimate, the CATE captures how the effect of ADHD varies across subpopulations:
\begin{equation}
    \tau(z) = E\left[Y \mid \text{do}(X = 1), Z = z\right] - E\left[Y \mid \text{do}(X = 0), Z = z\right]
    \label{eq:cate}
\end{equation}

This is a Layer~2 (interventional) quantity: it asks what the average treatment effect would be in a subpopulation defined by covariates $X = x$ under intervention, without requiring cross-world reasoning for the same individual. GRF estimates this function nonparametrically by recursively partitioning the covariate space to maximize treatment effect heterogeneity, producing individual-level CATE estimates $\hat{\tau}(x_i)$ with valid pointwise confidence intervals.

I examined heterogeneity along four dimensions: race/ethnicity, sex, socioeconomic status, and educational expectations. For each dimension, I reported group-level average CATEs with confidence intervals to identify which subgroups bear the largest or smallest ADHD penalties. To determine which covariates drove the most heterogeneity, I extracted variable importance measures from the random forest, which quantify each covariate's contribution to treatment effect variation across splits. This provided a data-driven ranking of the dimensions along which the ADHD penalty most varies, independent of any a priori assumptions about which subgroups are most affected.

To further explore how heterogeneity was jointly structured across dimensions, I constructed heatmaps of average CATEs cross-tabulated by race/ethnicity and educational expectations. These visualizations revealed whether certain combinations of social position and aspirational orientation concentrate or attenuate the ADHD penalty.

This step complemented the TV decomposition by shifting the analytic focus from channels (through what pathways does ADHD operate?) to populations (for whom is the ADHD effect strongest?).

\subsection{Counterfactual Direct Effect (ctf-DE) Estimation}

To address the third research question, I further examined heterogeneity in the direct causal effect specifically. While the CATE captures heterogeneity in the total effect of the treatment (combining all causal pathways), the ctf-DE captures heterogeneity only in the direct pathway that bypasses the measured mediators:
\begin{equation}
    \text{ctf-DE}(y \mid z) = E\left[Y_{x_1, W_{x_0}} \mid Z = z\right] - E\left[Y_{x_0, W_{x_0}} \mid Z = z\right]
    \label{eq:ctfde}
\end{equation}

This quantity asks: among students with characteristics $Z = z$, what would the ADHD STEM outcome gap be if ADHD-diagnosed students had the same mediator values they would have had without ADHD? It is a Layer~3 (counterfactual) quantity in the PCH, as it requires cross-world reasoning about the mediator value under a counterfactual treatment status.

I estimated ctf-DE using the one-step debiased (OSD) estimator proposed by \citet{Plecko2025} with 10-fold cross-fitting. Nuisance parameters, including outcome model and treatment propensity, were estimated using XGBoost to accommodate nonlinear relationships and interactions. The OSD estimator first obtains an initial estimate by plugging in predictions from the fitted nuisance models, then applies a one-step bias correction based on the influence function to remove first-order bias (the component of estimation error that scales linearly with nuisance model misspecification) and achieve semiparametric efficiency.

I also computed ctf-DEs across subgroups defined by race/ethnicity, socioeconomic status, and educational expectations to test whether the direct mechanism linking ADHD to STEM outcomes varied across students' social positions. If ctf-DEs differed substantially across subgroups, this would suggest that the unmediated ADHD penalty was shaped by students' social identities rather than operating uniformly.

\subsection{Sensitivity and Robustness}

Bias in causal inference arises from two sources: how strongly unmeasured factors affect the outcome (impact) and how unevenly these factors are distributed between treatment and control groups (imbalance). Because one cannot fully eliminate these bias sources due to unobserved confounders \citep{Pearl2021}, I evaluated the sensitivity of my estimates through two complementary analyses.

\noindent\textit{Omitted variable bias.} Using the \texttt{sensemakr} package proposed by \citet{Cinelli2020}, I quantified how strongly an unobserved confounder would need to influence both ADHD status and STEM outcomes to nullify the observed effects. Specifically, I computed robustness values (RV) that represent the minimum amount of variance in both the treatment and the outcome that an unmeasured confounder would need to explain to either reduce the causal estimate to zero ($\text{RV}_{q=1}$) or render it statistically non-significant ($\text{RV}_{\alpha=0.05}$). RV ranges from 0 to 1: the higher the value, the more robust the observed effect is to potential unobserved confounding \citep{Cinelli2020}. I benchmarked these values against the strongest observed confounders in the model, providing an interpretable scale for assessing vulnerability.

\noindent\textit{Overlap trimming.} The second analysis addressed violations of the positivity assumption---the requirement that all students have a nonzero probability of being in both groups. Students with extreme propensity scores contribute disproportionate variance. I progressively trimmed observations at the 1st through 5th percentiles of the propensity score distribution and re-estimated the full TV decomposition at each threshold. Stability across trimming levels indicated that results are not driven by students in regions of poor overlap, while substantial sensitivity would have flagged concerns about generalizability.

\subsection{Limitations}

This study has several limitations that should be considered when interpreting the results. First, the causal estimates depend on the correctness of the assumed causal graph. If the structural relationships among confounders, mediators, and outcomes are misspecified, the decomposition may not reflect the true causal structure. Second, the identification of causal effects requires sufficient overlap (common support) between ADHD and non-ADHD students across all confounder and mediator strata. Although approximately 10\% of the sample had an ADHD diagnosis and covariates spanned a wide range, certain subgroup intersections may have limited representation in one or both treatment conditions, making estimates in those regions sensitive to extrapolation. This concern was partially addressed by the overlap trimming analysis. Third, the nuisance parameters were estimated using machine learning methods (GRF, XGBoost), which are less interpretable than the OLS and logistic regression models traditionally used in educational research. Although they may lack some transparency, their use is standard practice in semiparametric causal inference, as imposing parametric assumptions on the complex relationships among ADHD status, demographic covariates, and STEM outcomes risks model misspecification bias.

\section{Findings}

\subsection{Descriptive Balance for ADHD and non-ADHD Students}

Table~\ref{tab:descriptive} presents descriptive statistics for the science identity and STEM GPA samples, stratified by ADHD status. SMDs reveal imbalance ($|\text{SMD}| > 0.10$) between ADHD and non-ADHD students. The largest imbalances appeared in sex (SMD $= -0.45$ for science identity; $-0.47$ for STEM GPA), with approximately 70\% of ADHD-diagnosed students being male compared to 48\% in the non-ADHD group. Substantial imbalances also emerged in educational expectations: students with ADHD were overrepresented among those expecting only a high school diploma (SMD $= 0.22$) and underrepresented among those expecting a PhD or professional degree (SMD $= -0.24$). Racial/ethnic composition differed as well, with ADHD students more likely to be White (SMD $= 0.18$) and less likely to be Asian (SMD $= -0.29$) or Hispanic/Latino (SMD $= -0.15$).

\begin{table}[!htb]
\centering
\caption{Descriptive Statistics and Standardized Mean Differences by ADHD Status}
\label{tab:descriptive}
\footnotesize
\renewcommand{\arraystretch}{1.1}
\begin{tabular}{l@{\hskip 6pt}c@{\hskip 6pt}c@{\hskip 6pt}c@{\hskip 8pt}c@{\hskip 6pt}c@{\hskip 6pt}c}
\toprule
 & \multicolumn{3}{c}{\textbf{Science Identity}} & \multicolumn{3}{c}{\textbf{STEM GPA}} \\
\cmidrule(lr){2-4} \cmidrule(lr){5-7}
\textbf{Variables} & \makecell{Non-ADHD\\(N=12,370)} & \makecell{ADHD\\(N=1,390)} & \textbf{SMD} & \makecell{Non-ADHD\\(N=13,030)} & \makecell{ADHD\\(N=1,570)} & \textbf{SMD} \\
\midrule
\multicolumn{7}{l}{\textit{Confounding Variables (Z)}} \\[3pt]
\textbf{Science Identity (9th)} & & & & & & \\
\quad Mean (SD) & 0.127 (0.99) & 0.067 (1.09) & $-$0.058 & 0.112 (0.99) & 0.080 (1.09) & $-$0.031 \\[3pt]
\textbf{Race/Ethnicity} & & & & & & \\
\quad White & 57.4\% & 65.9\% & \textbf{0.178} & 56.8\% & 65.0\% & \textbf{0.171} \\
\quad AI/AN/Pacific Isl. & 1.1\% & 1.1\% & 0.001 & 1.1\% & 1.4\% & 0.026 \\
\quad Asian & 8.2\% & 1.9\% & \textbf{$-$0.287} & 8.1\% & 2.0\% & \textbf{$-$0.280} \\
\quad Black & 9.2\% & 8.9\% & $-$0.010 & 9.4\% & 9.3\% & $-$0.002 \\
\quad Hispanic/Latino & 15.8\% & 10.7\% & \textbf{$-$0.151} & 16.2\% & 10.9\% & \textbf{$-$0.154} \\
\quad Multiracial & 8.3\% & 11.3\% & \textbf{0.100} & 8.4\% & 11.2\% & 0.093 \\[3pt]
\textbf{Sex} & & & & & & \\
\quad Male & 47.6\% & 69.2\% & \textbf{0.448} & 47.7\% & 70.3\% & \textbf{0.472} \\
\quad Female & 52.4\% & 30.8\% & \textbf{$-$0.448} & 52.3\% & 29.7\% & \textbf{$-$0.472} \\[3pt]
\textbf{Family SES Quintile} & & & & & & \\
\quad Q1 (Lowest) & 15.4\% & 16.8\% & 0.037 & 16.5\% & 19.2\% & 0.070 \\
\quad Q2 & 16.0\% & 18.4\% & 0.063 & 15.9\% & 19.6\% & 0.099 \\
\quad Q3 & 16.9\% & 18.9\% & 0.053 & 17.1\% & 17.8\% & 0.018 \\
\quad Q4 & 19.5\% & 17.9\% & $-$0.039 & 19.2\% & 17.3\% & $-$0.049 \\
\quad Q5 (Highest) & 32.1\% & 27.8\% & $-$0.093 & 31.3\% & 26.0\% & \textbf{$-$0.116} \\[3pt]
\textbf{Highest Educ.\ Expectation} & & & & & & \\
\quad Don't know & 19.4\% & 26.7\% & \textbf{0.177} & 19.5\% & 27.4\% & \textbf{0.188} \\
\quad High School & 9.3\% & 16.4\% & \textbf{0.215} & 9.8\% & 17.6\% & \textbf{0.229} \\
\quad Associate's & 5.4\% & 8.4\% & \textbf{0.116} & 5.5\% & 8.2\% & \textbf{0.107} \\
\quad Bachelor's & 17.9\% & 16.3\% & $-$0.043 & 17.8\% & 15.6\% & $-$0.060 \\
\quad Master's & 23.4\% & 17.2\% & \textbf{$-$0.153} & 23.2\% & 16.3\% & \textbf{$-$0.173} \\
\quad PhD/Professional & 24.6\% & 14.8\% & \textbf{$-$0.247} & 24.2\% & 14.8\% & \textbf{$-$0.238} \\[3pt]
\midrule
\multicolumn{7}{l}{\textit{Mediating Variables (W)}} \\[3pt]
\textbf{Science Club} & & & & & & \\
\quad Yes & 7.2\% & 4.6\% & \textbf{$-$0.109} & 7.2\% & 4.5\% & \textbf{$-$0.113} \\[3pt]
\textbf{Taking Algebra II} & & & & & & \\
\quad Yes & 38.9\% & 44.5\% & \textbf{0.114} & 39.8\% & 42.9\% & 0.064 \\[3pt]
\textbf{Taking Geometry} & & & & & & \\
\quad Yes & 13.0\% & 25.0\% & \textbf{0.309} & 13.4\% & 26.9\% & \textbf{0.341} \\[3pt]
\textbf{Taking Pre-calculus} & & & & & & \\
\quad Yes & 26.8\% & 10.5\% & \textbf{$-$0.427} & 25.4\% & 9.5\% & \textbf{$-$0.429} \\
\bottomrule
\end{tabular}
\end{table}

Among mediators, the most notable imbalance appeared in geometry enrollment (SMD $= 0.31$ for science identity; $0.34$ for STEM GPA), indicating that ADHD students were substantially more likely to be taking geometry in 11th grade. Science club participation also differed (SMD $= -0.11$), with ADHD students less likely to participate.

\subsection{Total Variation Decomposition}

\begin{figure}[H]
    \centering
    \includegraphics[width=0.65\textwidth]{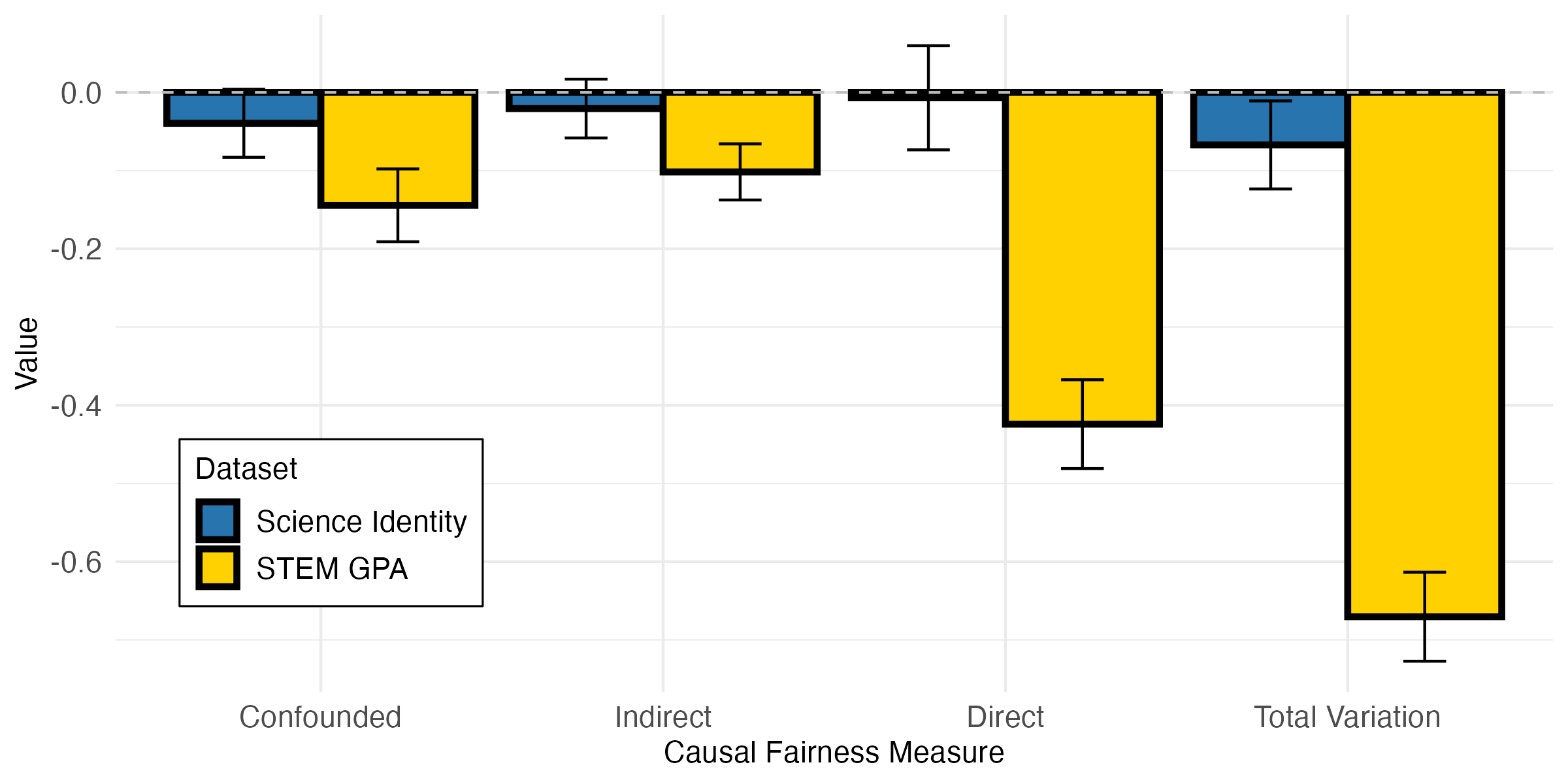}
    \caption{Total variation decomposition of the ADHD--STEM outcome disparity}
    \label{fig:tvdecomp}
\end{figure}

The results quantifying confounded, indirect, and direct effects on both outcomes are shown in Figure~\ref{fig:tvdecomp}. The TV measure took the same negative sign on both outcomes but differed dramatically in magnitude. When applying the causal decomposition, we obtain:
\begin{align}
    \text{Science Identity:}\quad & E[Y \mid \text{ADHD}] - E[Y \mid \text{non-ADHD}] = -0.068 \notag \\
    &= \underbrace{-0.040}_{\text{confounded}} + \underbrace{(-0.021)}_{\text{indirect}} + \underbrace{(-0.007)}_{\text{direct}} \label{eq:sciid} \\[20pt]
    \text{STEM GPA:}\quad & E[Y \mid \text{ADHD}] - E[Y \mid \text{non-ADHD}] = -0.670 \notag \\
    &= \underbrace{-0.144}_{\text{confounded}} + \underbrace{(-0.102)}_{\text{indirect}} + \underbrace{(-0.424)}_{\text{direct}} \label{eq:stemgpa}
\end{align}

The two outcomes told different stories in their causal structure, despite sharing the same sign. For STEM GPA, the direct causal pathway ($x$-DE $= -0.424$, SD $= 0.029$) was the dominant component, accounting for 63.3\% of the total gap. This means that for two students with comparable characteristics (demographic background, educational engagement) who differ only in ADHD diagnosis status, the ADHD-diagnosed student would be expected, on average, to score 0.42 SD lower in STEM GPA. The confounded causal effect ($x$-SE $= -0.144$, SD $= 0.024$) accounted for 21.6\% of the total variation, transmitted through demographic variables such as sex, SES, race/ethnicity, and educational expectations. This indicated that ADHD-diagnosed students were drawn from demographic strata that independently predict lower STEM GPA. The indirect causal pathway ($x$-IE $= -0.102$, SD $= 0.018$) accounts for 15.2\% of the total variation. Variables that transmitted this indirect effect included advanced math enrollment and science club participation, indicating that ADHD shaped educational engagement, which in turn reduced STEM GPA outcome. All of the effects in the STEM GPA outcome were statistically significant.

For science identity, the total variation was much smaller (TV $= -0.067$, SD $= 0.029$) than for STEM GPA. The confounded effect was the dominant component ($x$-SE $= -0.040$), accounting for 58.9\% of the total variation, meaning that most of the modest gap was attributable to demographic composition rather than any causal effect of ADHD. The direct effect was negligible ($x$-DE $= -0.007$, SD $= 0.034$), indicating that ADHD status did not directly affect how students identified with science. Although the total variation was statistically significant, none of the individual decomposition components reached significance. Due to the small variation in science identity outcomes, the remaining findings focus primarily on STEM GPA.

\subsection{Heterogeneity in the Total Causal Effect}

The average treatment effect (ATE) estimated via GRF further confirmed this variation decomposition pattern. For STEM GPA, the ATE was $-0.530$ (SD $= 0.031$, 95\% CI: [$-0.59$, $-0.47$]), indicating a substantial and statistically significant causal penalty. For science identity, the ATE was $-0.026$ (SD $= 0.034$, 95\% CI: [$-0.09$, $0.04$]), which was not statistically significant.

Variable importance measures from the GRF indicated that baseline science identity, race/ethnicity, and educational expectations were the primary drivers of treatment effect heterogeneity for both outcomes, while sex contributed minimally (0.052 for science identity, 0.027 for STEM GPA). The negligible role of sex in driving heterogeneity is notable given its large baseline imbalance, suggesting that while ADHD diagnosis is strongly gendered, the effect of ADHD on STEM outcomes does not differ substantially between males and females. Appendix~D presents full subgroup CATEs by sex, race/ethnicity, educational expectations, and SES quintile. For STEM GPA, the ADHD penalty was consistent across sexes (males: $-0.504$; females: $-0.512$) but varied along other dimensions. Multiracial students bore the largest penalty ($-0.609$), followed by Asian ($-0.528$) and White ($-0.521$) students, while Black ($-0.431$) and Hispanic/Latino ($-0.442$) students showed somewhat smaller effects. The penalty increased monotonically with both educational expectations (from $-0.333$ for high school to $-0.567$ for PhD/professional) and SES (from $-0.405$ in Q1 to $-0.540$ in Q5). For science identity, CATEs were uniformly small, with one exception that multiracial students showed the largest subgroup effect ($-0.112$).

\begin{figure}[H]
    \centering
    \includegraphics[width=1.1\textwidth]{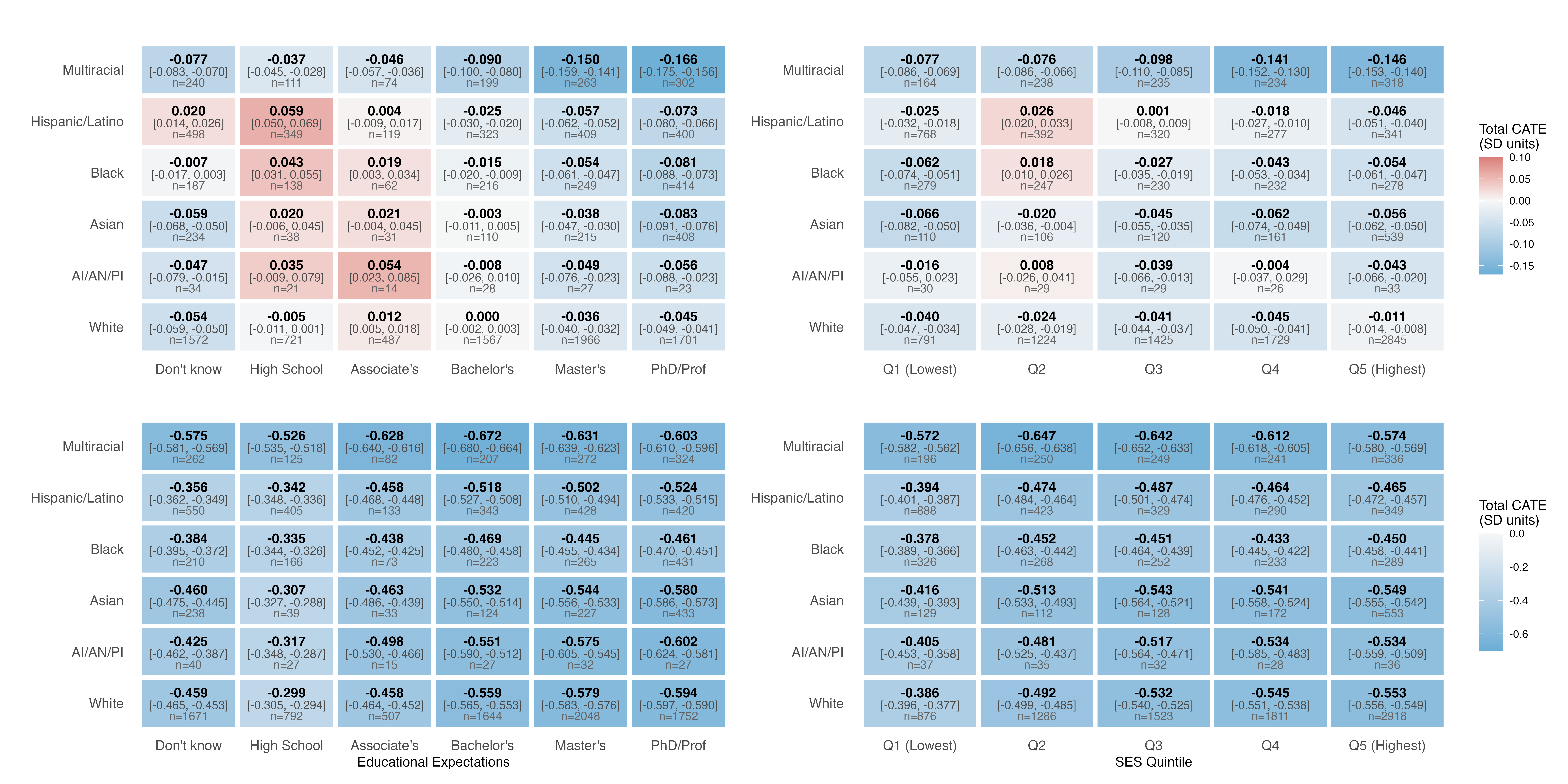}
    \caption{Heatmaps of CATE of ADHD on science identity (top) and STEM GPA (bottom)}
    \label{fig:cateheatmap}
\end{figure}

I further examined cross-tabulated CATEs by race/ethnicity and highest educational expectation, and by race and SES. The heatmaps was presented in Figure ~\ref{fig:cateheatmap}. For STEM GPA, the race-by-educational expectations heatmap showed that the most severe penalties concentrated among multiracial students with Bachelor's-level expectations ($-0.672$), while the smallest penalties appeared among students expecting only a high school diploma across most racial groups ($-0.299$ to $-0.342$). The race-by-SES heatmaps revealed that ADHD penalty among White students increased from $-0.386$ (lowest SES) to $-0.553$ (highest SES). For science identity, most cells remained near zero, but Hispanic/Latino students showed a notable exception: those expecting a high school diploma ($+0.059$) exhibited small positive CATEs, suggesting that ADHD diagnosis may carry a slight protective effect on science identity for lower-aspiration Hispanic/Latino students. 

\subsection{Heterogeneity in the Direct Causal Effect}

\begin{figure}[htbp]
    \centering
    \includegraphics[width=1.1\textwidth]{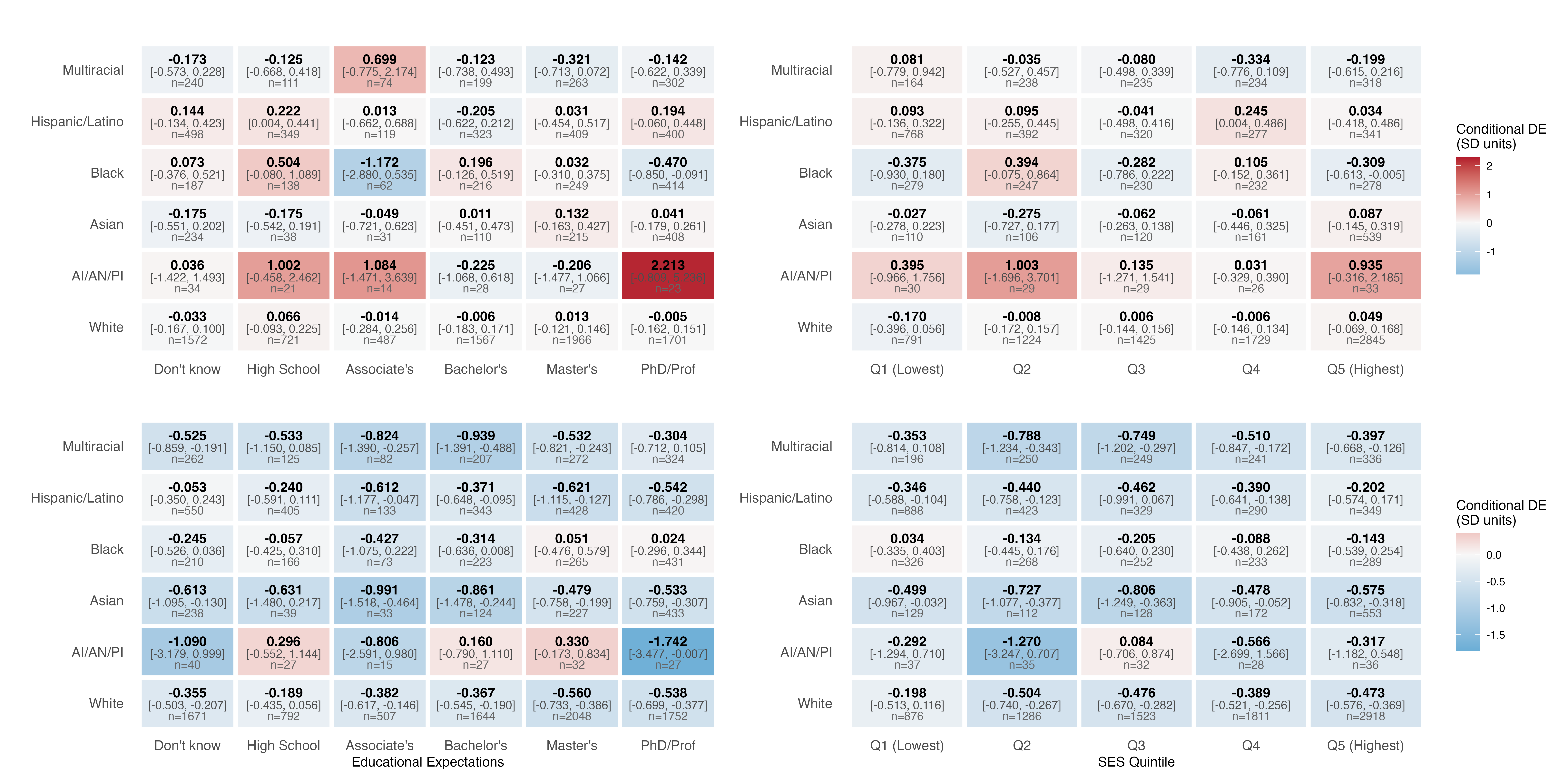}
    \caption{Heatmaps of direct effect heterogeneity on science identity (top) and STEM GPA (bottom)}
    \label{fig:cdeheatmap}
\end{figure}

Figure~\ref{fig:cdeheatmap} presents heatmaps of ctf-DEs across subgroups defined by race/ethnicity crossed with educational expectations and SES quintile. For STEM GPA, the ctf-DE heatmaps revealed that the unmediated ADHD penalty varied substantially by race. Asian students showed consistently large direct effects across most cells (e.g., $-0.991$ for Associate's expectations; $-0.806$ for Q3 SES), while Black students showed notably small or even slightly positive effects in several cells ($0.051$ for Master's; $0.034$ for Q1 SES), suggesting the unmediated penalty was attenuated for this group. White students showed moderate direct effects that increased at higher expectation and SES levels ($-0.560$ for Master's; $-0.504$ for Q2 SES), and multiracial students showed large penalties at mid-to-high expectations ($-0.939$ for Bachelor's). These patterns suggest that the unmediated ADHD penalty is itself structured by students' social positions, pointing toward differentiated rather than uniform processes of disadvantage. Estimates for AI/AN/PI students showed extreme values with very wide confidence intervals due to small cell sizes (often $n < 30$) and should be interpreted with caution. For science identity, the ctf-DEs are generally close to zero, which is consistent with the negligible population-level direct effect.

\subsection{Sensitivity and Robustness}

Regarding sensitivity to omitted variable bias, the STEM GPA estimate showed considerable robustness. The $\text{RV}_{q=1}$ was 14.19\%, indicating that an unobserved confounder would need to explain at least 14.19\% of the residual variance in both ADHD status and STEM GPA to reduce the point estimate to zero. Similarly, the $\text{RV}_{\alpha=0.05}$ was 12.79\%, meaning a confounder would need to explain at least 12.79\% of the residual variance in both ADHD status and STEM GPA to render the effect statistically insignificant at the $\alpha = 0.05$ level. To put these thresholds in context, the observed covariate with the strongest association with treatment assignment was sex ($R^2_{dz.x} = 1.72\%$), and the strongest predictor of the outcome was SES ($R^2_{yz.dx} = 7.75\%$), both well below the robustness thresholds. An unobserved confounder would therefore need to be substantially stronger than any observed covariate to overturn the STEM GPA finding. For science identity, both $\text{RV}_{q=1}$ and $\text{RV}_{\alpha=0.05}$ were near zero, indicating that even a trivially small unobserved confounder could nullify the estimated effect. This was consistent with the near-zero and non-significant direct effect reported above.

\begin{figure}[H]
    \centering
    \includegraphics[width=0.59\textwidth]{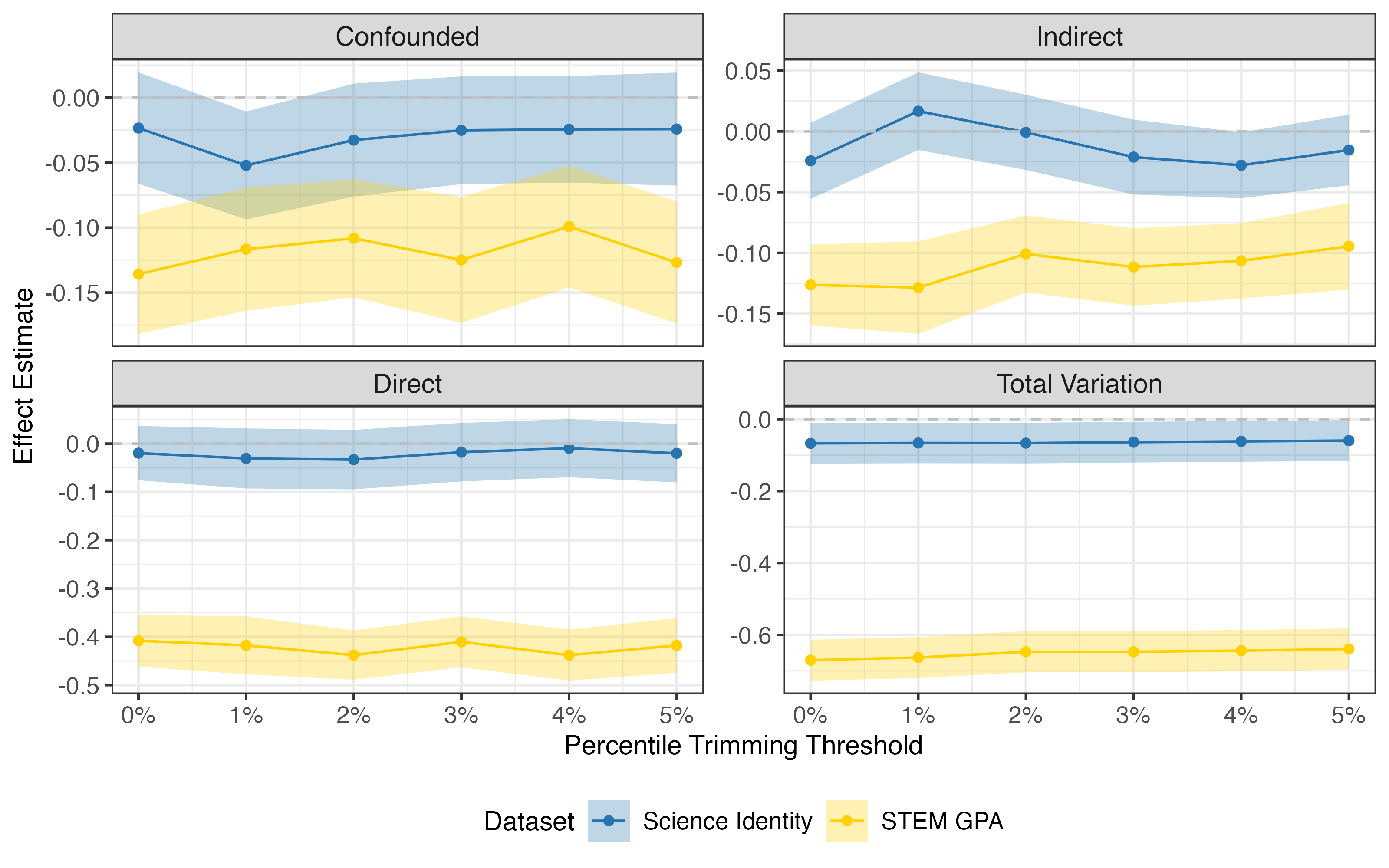}
    \caption{Propensity Score Overlap Trimming Sensitivity Analysis for TV Decomposition with 95\% confidence intervals}
    \label{fig:overlap}
\end{figure}

In addition, Figure~\ref{fig:overlap} shows the TV decomposition re-estimated at progressive trimming thresholds (1st--5th percentiles). For STEM GPA, all four components (TV, DE, IE, SE) remained stable across trimming levels, with point estimates varying by less than 0.05 SD. For science identity, the estimates were similarly stable near zero. The stability of the decomposition under increasingly restrictive overlap conditions indicated that the causal findings were not driven by students with extreme propensity scores.

\section{Conclusion and Discussion}

This study applied the CFA \citep{Plecko2024} to decompose the observed ADHD--STEM outcome disparity into confounded, indirect, and direct components, and examined heterogeneity in both the total and direct effects across student demographic subgroups. The analysis revealed three main findings.

First, the ADHD penalty varied across STEM outcomes. The TV decomposition revealed a substantial disparity in STEM GPA (TV $= -0.670$) but only a negligible gap in science identity (TV $= -0.067$). For STEM GPA, all three decomposition components were statistically significant, with the direct effect accounting for 63.3\% of the total gap. This indicates that the majority of the penalty operated through channels not captured by the measured mediators, potentially reflecting institutional responses, special education practices, or stigma associated with the ADHD label, though this estimate may also partially absorb indirect pathways through unmeasured mediators. For science identity, the total gap was small and none of the individual decomposition components reached statistical significance. This divergence indicates that ADHD disrupted how students performed in STEM but not how they identified with science. In other words, ADHD-diagnosed students may retain the motivational and identity resources needed to persist in STEM despite facing performance barriers. This finding extends \citet{Shifrer2021}, who documented that medicated ADHD youth, despite showing the highest STEM achievement among disability groups, still lagged behind non-disabled peers in both achievement and science identity. However, their regression-based approach could not determine whether these gaps reflected causal effects of ADHD or preexisting demographic differences. The present study's causal decomposition clarifies that the STEM GPA gap was driven primarily by direct and indirect effects of ADHD, whereas the science identity gap was small, non-significant at the component level, and largely attributable to confounding.

Second, the ADHD penalty was most severe among students positioned for academic success. The CATE analysis revealed that the ADHD penalty on STEM GPA increased monotonically with educational expectations (from $-0.333$ for high school to $-0.567$ for PhD/professional) and SES (from $-0.405$ in Q1 to $-0.540$ in Q5). These findings challenge the assumption that disadvantaged students with ADHD are the most affected; instead, the penalty was most acute among students with high aspirations and socioeconomic advantage. One potential interpretation is that high-expectation, high-SES educational environments impose greater demands on executive function and self-regulation (capacities impaired by ADHD), which amplify the consequences of an ADHD diagnosis in these contexts.

Third, the direct effect differed by race. The ctf-DE analysis revealed that the unmediated ADHD penalty varied substantially by race/ethnicity. Asian students showed the largest direct effects across several subgroup intersections, while Black students showed notably small or slightly positive effects. These patterns indicate that the direct ADHD penalty was not uniform but was itself shaped by students' racial/ethnic positions.

Moreover, this study demonstrates that the CFA framework offers a valuable alternative to conventional regression-based methods in educational research. Traditional regression-based approaches, even when applying propensity score weighting and matching, typically estimate a single aggregate treatment effect without distinguishing whether the disparity operates through direct, indirect, or spurious pathways. The CFA framework moves beyond this by enabling researchers to quantify how much of a gap is causal versus confounded, whether it flows through mediating pathways or bypasses them, and whether its causal structure varies across social groups. This framework and the current study provide a replicable template for other educational equity questions involving a protected attribute (such as gender, race/ethnicity) linked to outcomes through multiple effect channels.

Building on the current study, several directions for future research warrant consideration. The direct effect accounted for nearly two-thirds of the STEM GPA gap. Incorporating additional mediators, such as teacher perceptions, targeted special education practices, or extracurricular activities would allow future studies to further decompose the direct pathway. The racial variation in both CATE and ctf-DE also warrants deeper investigation. Qualitative or mixed-methods research could help explain how the direct penalty is attenuated for some racial groups but amplified for others. Such work might examine how institutional responses to ADHD vary across school contexts serving different racial/ethnic populations, and how family background shapes access to accommodations, special education, and diagnostic resources. Finally, the finding that ADHD substantially penalized STEM GPA but did not directly affect science identity warrants longitudinal investigation. It remains unclear whether this identity resilience persists into postsecondary education, or whether sustained performance barriers eventually erode it. Tracking the STEM trajectories of ADHD-diagnosed students beyond high school would clarify whether the science identity resilience observed here translates into durable STEM persistence.

\bigskip
\noindent\textbf{Code Availability.} The code repository is available at: \url{https://github.com/AliceAii/ADHD_STEM-Outcomes_Causal-Fairness-Analysis}.

\newpage
\appendix

\section*{Appendix A. Analytic pipeline}
\addcontentsline{toc}{section}{Appendix A}

\begin{figure}[H]
    \centering
    \includegraphics[width=1.1\textwidth]{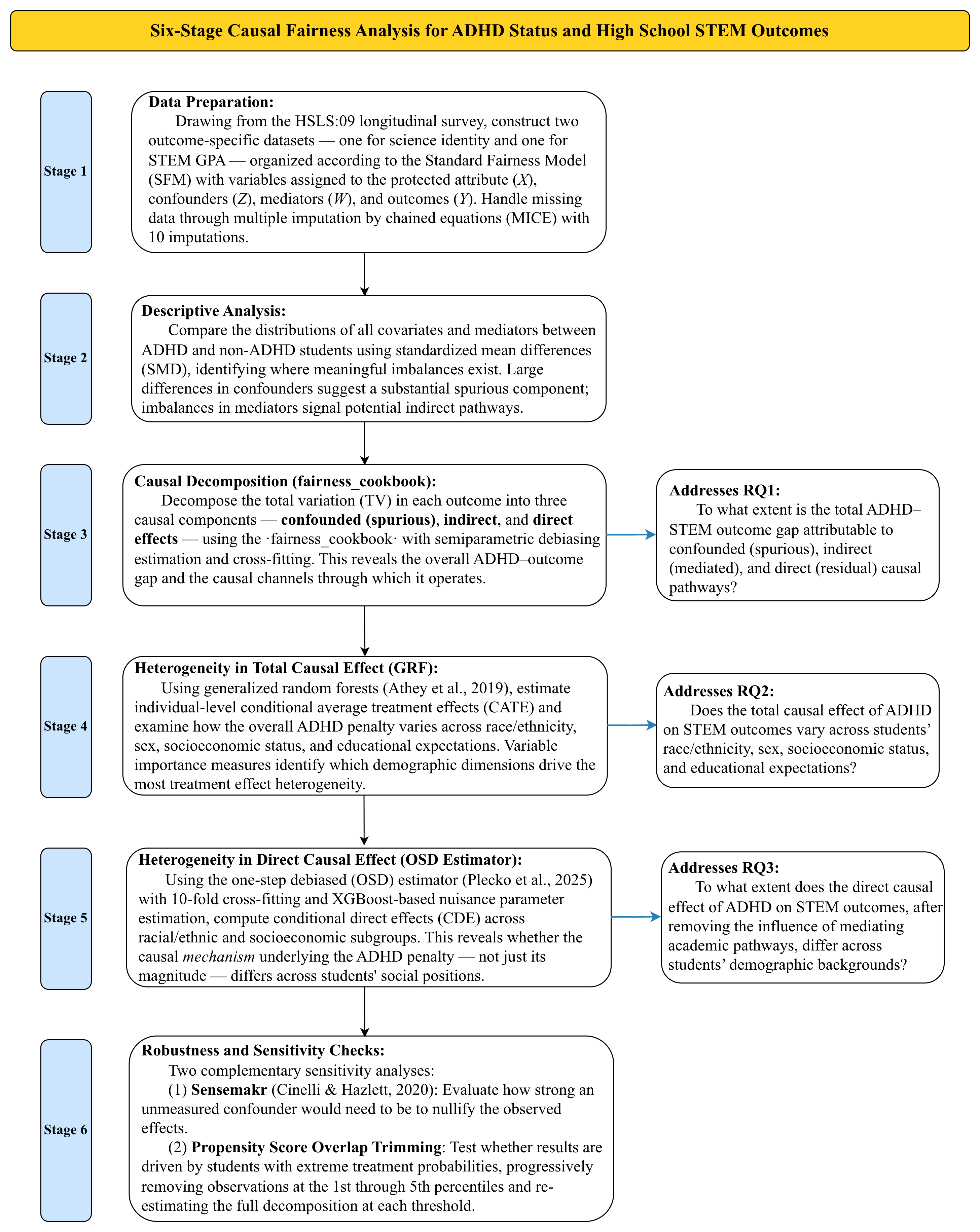}
\end{figure}

\newpage
\section*{Appendix B. Variable definitions and coding details}
\addcontentsline{toc}{section}{Appendix B}

\begin{table}[htbp]
\centering
\footnotesize
\renewcommand{\arraystretch}{1.2}
\begin{tabular}{p{4.5cm}p{2.2cm}p{7cm}}
\toprule
\textbf{Variables} & \textbf{HSLS Variable} & \textbf{Coding Description} \\
\midrule
\multicolumn{3}{l}{\textit{Protected Attribute (X)}} \\
ADHD diagnosis & ADHD & Factor: 1 = Yes; 0 = No (ref) \\
\midrule
\multicolumn{3}{l}{\textit{Confounding Variables (Z)}} \\
\textbf{Student Background} & & \\
Race/ethnicity composite & X1RACE & Factor: White (ref), Amer.\ Indian/Alaska Native/Pacific Islander, Asian, Black, Hispanic/Latino, Multiracial \\
Student's sex & X1SEX & Factor: 1 = Female; 0 = Male (ref) \\
Socioeconomic status quintile & X1SESQ5 & Factor: From 5 = Fifth quintile (highest) to 1 = First quintile (lowest) (ref) \\
\textbf{Competence \& Perception} & & \\
9th-grade science identity scale & X1SCIID & NCES standardized latent construct. Standardized to a mean of 0 and standard deviation of 1. \\
9th-grade highest educational expectation & X1STUEDEXPCT & Factor: Don't know (ref), High School, Associate's, Bachelor's, Master's, PhD/Professional \\
\midrule
\multicolumn{3}{l}{\textit{Mediating Variables (W)}} \\
\textbf{Formal and Informal STEM Learning Experiences (11th grade)} & & \\
Participated in a science club or science study group between 9th and 11th grade & S2SCLUB & Factor: 1 = Yes; 0 = No (ref) \\
Taking Algebra II in spring 2012 & S2ALG2M12 & Factor: 1 = Yes; 0 = No (ref) \\
Taking Geometry in spring 2012 & S2GEOM12 & Factor: 1 = Yes; 0 = No (ref) \\
Taking Pre-calculus or Analysis and Functions in spring 2012 & S2PRECALC12 & Factor: 1 = Yes; 0 = No (ref) \\
\midrule
\multicolumn{3}{l}{\textit{Outcome Variables (Y)}} \\
11th-grade science identity scale & X2SCIID & NCES standardized latent construct. Continuous, standardized to a mean of 0 and standard deviation of 1. \\
High school STEM GPA & X3TGPASTEM & Continuous, 0--4 scale, standardized to a mean of 0 and standard deviation of 1. \\
\bottomrule
\end{tabular}
\end{table}

\newpage
\section*{Appendix C. Counterfactual definitions and identification of each component}
\addcontentsline{toc}{section}{Appendix C}

\subsection*{C.1 Counterfactual Definitions}

\tikzset{
    dot/.style={circle, draw=black, thick, minimum size=3.5mm, inner sep=0pt},
    reddot/.style={dot, fill=red},
    bluedot/.style={dot, fill=cyan!70!blue},
    arr/.style={->, thick, color=black, >={Stealth[length=2mm, width=1.5mm]}},
    dasharr/.style={<->, dashed, thick, color=black, >={Stealth[length=2mm, width=1.5mm]}},
    redarr/.style={->, line width=1.6pt, color=red, >={Stealth[length=2mm, width=1.5mm]}},
    cyanarr/.style={->, line width=1.6pt, color=cyan, >={Stealth[length=2mm, width=1.5mm]}},
    reddash/.style={<->, dashed, line width=1.2pt, color=red, >={Stealth[length=2mm, width=1.5mm]}},
    cyandash/.style={<->, dashed, line width=1.2pt, color=cyan!80!blue, >={Stealth[length=2mm, width=1.5mm]}},
    ominus/.style={line width=3pt, color=orange!90!yellow},
}

%% ==========================================================
%% 1. DIRECT EFFECT
%% ==========================================================
\textbf{(i) Direct effect ($x$-DE).} The $x$-specific direct effect measures the change in $Y$ when $X$ is shifted from $x_0$ to $x_1$ along $X \to Y$, while keeping the mediators $W$ at their natural level under $X = x_0$:
\begin{equation}
x\text{-}\mathbf{DE}_{x_0,x_1}(y \mid x_0) = E\!\left[Y_{x_1, W_{x_0}} - Y_{x_0} \;\middle|\; X = x_0\right] \tag{C1}
\end{equation}
This captures the residual disparity that would remain if ADHD and non-ADHD students had identical educational engagement profiles and background characteristics.

\smallskip
\begin{center}
\begin{tikzpicture}[scale=0.7, transform shape]
    % --- LEFT DAG ---
    \node[reddot] (X0) at (0,0) {}; \node[left=1.5mm of X0, font=\small] {$X\!=\!x_0$};
    \node[bluedot, label={[font=\small]above:$Z$}] (Z) at (1.5, 1.2) {};
    \node[bluedot, label={[font=\small]below:$W$}] (W) at (1.5, -1.2) {};
    \node[bluedot, label={[font=\small]right:$Y$}] (Y) at (3, 0) {};
    \node[reddot] (X1) at (1.5, 0) {}; \node[above=1mm of X1, font=\small] {$X\!=\!x_1$};
    \draw[dasharr] (X0) -- (Z);
    \draw[arr] (Z) -- (Y);
    \draw[arr] (X0) -- (W);
    \draw[arr] (W) -- (Y);
    \draw[redarr] (X1) -- (Y);
    \node[font=\normalsize] at (1.5, -2) {$Y_{x_1, W_{x_0}} \mid X = x_0$};

    % --- MINUS ---
    \draw[ominus] (4.6, 0) -- (5.2, 0);

    % --- RIGHT DAG ---
    \begin{scope}[xshift=7cm]
        \node[reddot] (X0) at (0,0) {}; \node[left=1.5mm of X0, font=\small] {$X\!=\!x_0$};
        \node[bluedot, label={[font=\small]above:$Z$}] (Z) at (1.5, 1.2) {};
        \node[bluedot, label={[font=\small]below:$W$}] (W) at (1.5, -1.2) {};
        \node[bluedot, label={[font=\small]right:$Y$}] (Y) at (3, 0) {};
        \draw[dasharr] (X0) -- (Z);
        \draw[arr] (Z) -- (Y);
        \draw[arr] (X0) -- (W);
        \draw[arr] (W) -- (Y);
        \draw[cyanarr] (X0) -- (Y);
        \node[font=\normalsize] at (1.5, -2) {$Y_{x_0, W_{x_0}} \mid X = x_0$};
    \end{scope}
\end{tikzpicture}
\end{center}

\medskip

%% ==========================================================
%% 2. INDIRECT EFFECT
%% ==========================================================
\textbf{(ii) Indirect effect ($x$-IE).} The $x$-specific indirect effect measures the change in $Y$ transmitted through $W$ when $X$ is shifted from $x_1$ to $x_0$ along $X \to W \to Y$, while holding $X$ fixed at $x_1$ along the direct path:
\begin{equation}
x\text{-}\mathbf{IE}_{x_1,x_0}(y \mid x_0) = E\!\left[Y_{x_1, W_{x_0}} - Y_{x_1} \;\middle|\; X = x_0\right] \tag{C2}
\end{equation}
This quantity captures how much of the disparity is transmitted because ADHD affects educational engagement (e.g., science club participation, STEM class completion), which in turn shapes STEM outcomes.

\smallskip
\begin{center}
\begin{tikzpicture}[scale=0.7, transform shape]
    % --- LEFT DAG ---
    \node[reddot] (X0) at (0,0) {}; \node[left=1.5mm of X0, font=\small] {$X\!=\!x_0$};
    \node[bluedot, label={[font=\small]above:$Z$}] (Z) at (1.5, 1.2) {};
    \node[bluedot, label={[font=\small]below:$W$}] (W) at (1.5, -1.2) {};
    \node[bluedot, label={[font=\small]right:$Y$}] (Y) at (3, 0) {};
    \node[reddot] (X1) at (0.55, -0.95) {}; \node[left=1mm of X1, font=\small] {$X\!=\!x_1$};
    \draw[dasharr] (X0) -- (Z);
    \draw[arr] (Z) -- (Y);
    \draw[arr] (X0) -- (Y);
    \draw[redarr] (X1) -- (W);
    \draw[redarr] (W) -- (Y);
    \node[font=\normalsize] at (1.5, -2) {$Y_{x_0, W_{x_1}} \mid X = x_0$};

    % --- MINUS ---
    \draw[ominus] (4.6, 0) -- (5.2, 0);

    % --- RIGHT DAG ---
    \begin{scope}[xshift=7cm]
        \node[reddot] (X0) at (0,0) {}; \node[left=1.5mm of X0, font=\small] {$X\!=\!x_0$};
        \node[bluedot, label={[font=\small]above:$Z$}] (Z) at (1.5, 1.2) {};
        \node[bluedot, label={[font=\small]below:$W$}] (W) at (1.5, -1.2) {};
        \node[bluedot, label={[font=\small]right:$Y$}] (Y) at (3, 0) {};
        \draw[dasharr] (X0) -- (Z);
        \draw[arr] (Z) -- (Y);
        \draw[arr] (X0) -- (Y);
        \draw[cyanarr] (X0) -- (W);
        \draw[cyanarr] (W) -- (Y);
        \node[font=\normalsize] at (1.5, -2) {$Y_{x_0, W_{x_0}} \mid X = x_0$};
    \end{scope}
\end{tikzpicture}
\end{center}

\medskip

%% ==========================================================
%% 3. SPURIOUS EFFECT
%% ==========================================================
\textbf{(iii) Spurious/Confounded effect ($x$-SE).} The $x$-specific spurious effect isolates variation through non-causal pathways $X \leftrightarrow Z \to Y$, comparing the counterfactual outcome under the same intervention across populations differing only in natural group membership:
\begin{equation}
x\text{-}\mathbf{SE}_{x_1,x_0}(y) = E\!\left[Y_{x_1} \;\middle|\; X = x_0\right] - E\!\left[Y_{x_1} \;\middle|\; X = x_1\right] \tag{C3}
\end{equation}
This captures disparities attributable to demographic composition. For instance, if ADHD diagnosis rates vary by SES, and SES independently affects STEM outcomes, then even under identical treatment the two groups would exhibit different outcomes because they are drawn from different confounder strata.

\smallskip
\begin{center}
\begin{tikzpicture}[scale=0.7, transform shape]
    % --- LEFT DAG ---
    \node[reddot, label={[font=\small]above:{$X\!=\!x_1$}}] (X1) at (0,0) {};
    \node[bluedot, label={[font=\small]above:$Z$}] (Z) at (0, 1.4) {};
    \node[bluedot, label={[font=\small]below:$W$}] (W) at (0, -1.4) {};
    \node[bluedot, label={[font=\small]right:$Y$}] (Y) at (2, 0) {};
    \node[reddot, label={[font=\small]below:$X\!=\!x_0$}] (X0) at (-1.6, 0.4) {};
    \draw[arr] (X1) -- (Y);
    \draw[arr] (X1) -- (W);
    \draw[arr] (Z) -- (Y);
    \draw[arr] (W) -- (Y);
    \draw[reddash] (X0) -- (Z);
    \node[font=\normalsize] at (0, -2.2) {$Y_{x_1} \mid X = x_0$};

    % --- MINUS ---
    \draw[ominus] (4.1, 0) -- (4.7, 0);

    % --- RIGHT DAG ---
    \begin{scope}[xshift=7cm]
        \node[reddot] (X1) at (0,0) {}; \node[below=1.5mm of X1, font=\small] {$X\!=\!x_1$};
        \node[bluedot, label={[font=\small]above:$Z$}] (Z) at (1.5, 1.2) {};
        \node[bluedot, label={[font=\small]below:$W$}] (W) at (1.5, -1.2) {};
        \node[bluedot, label={[font=\small]right:$Y$}] (Y) at (3, 0) {};
        \draw[cyandash] (X1) -- (Z);
        \draw[arr] (Z) -- (Y);
        \draw[arr] (X1) -- (W);
        \draw[arr] (X1) -- (Y);
        \draw[arr] (W) -- (Y);
        \node[font=\normalsize] at (1.5, -2) {$Y_{x_1} \mid X = x_1$};
    \end{scope}
\end{tikzpicture}
\end{center}

\subsection*{C.2 Identification of $x$-DE under the SFM}

The central identification challenge is expressing the nested counterfactual $P(y_{x, W_{x'}} \mid x')$ in terms of observable quantities. We focus on this term since $E[Y_{x_0} \mid X = x_0] = E[Y \mid X = x_0]$ by consistency.

First, un-nest via the law of total probability and introduce $Z$:
\begin{equation}
P(y_{x, W_{x'}} \mid x') = \sum_{w,z} P(y_{xw},\; w_{x'} \mid z, x') \cdot P(z \mid x') \tag{C4}
\end{equation}

The key step is factorizing the cross-world joint $P(y_{xw}, w_{x'} \mid z, x')$. Constructing the Ancestral Multi-World Network \citep[AMWN;][]{Plecko2024} for the SFM yields ancestor sets $An(Y_{xw}) = \{Y_{xw}, Z\}$ and $An(W_{x'}) = \{W_{x'}, Z\}$, with $Z$ as a shared node forming a fork $W_{x'} \leftarrow Z \rightarrow Y_{xw}$. Conditioning on $Z$ d-separates $Y_{xw}$ and $W_{x'}$, giving:
\begin{equation}
P(y_{xw}, w_{x'} \mid z, x') = P(y_{xw} \mid z, x') \cdot P(w_{x'} \mid z, x') \tag{C5}
\end{equation}

Each factor reduces to observational quantities through the identification assumptions:

\begin{itemize}
    \item \textit{Consistency:} Since we condition on $X = x'$, which matches the intervention $\text{do}(X := x')$, we have $P(w_{x'} \mid z, x') = P(w \mid z, x')$.
    \item \textit{Conditional ignorability:} From the AMWN, $Y_{xw}$ depends only on $Z$ and the exogenous variable $U_Y$. Given $Z = z$, $Y_{xw}$ is independent of $X$, so $P(y_{xw} \mid z, x') = P(y_{xw} \mid z)$.
    \item \textit{Back-door criterion:} In the SFM, there is no hidden confounding between $(X, W)$ and $Y$ beyond $Z$, so $P(y_{xw} \mid z) = P(y \mid x, w, z)$.
\end{itemize}

Substituting back into (C4):
\begin{equation}
\boxed{P(y_{x, W_{x'}} \mid x') = \sum_{w,z} P(y \mid x, w, z) \cdot P(w \mid x', z) \cdot P(z \mid x')} \tag{C6}
\end{equation}

Every term on the right-hand side is estimable from observational data: $P(y \mid x, w, z)$ is the conditional outcome distribution stratified on $X$, $W$, and $Z$; $P(w \mid x', z)$ is the mediator distribution among group $x'$ given confounders; and $P(z \mid x')$ is the confounder distribution in group $x'$. Therefore, the complete $x$-DE identification formula is (setting $x = x_1$, $x' = x_0$):
\begin{equation}
x\text{-DE}_{x_0,x_1}(y \mid x_0) = \sum_{w,z} E[Y \mid x_1, w, z] \cdot P(w \mid x_0, z) \cdot P(z \mid x_0) - E[Y \mid x_0] \tag{C7}
\end{equation}

\subsection*{C.4 Identification Results for $x$-IE and $x$-SE}

Analogous derivations yield the following identification formulas for the remaining two components:
\begin{equation}
x\text{-IE}_{x_1,x_0}(y \mid x_0) = \sum_{w,z} E[Y \mid x_0, w, z] \cdot \Big[P(w \mid x_0, z) - P(w \mid x_1, z)\Big] \cdot P(z \mid x_0) \tag{C8}
\end{equation}

\begin{equation}
x\text{-SE}_{x_1,x_0}(y) = \sum_{w,z} E[Y \mid x_0, w, z] \cdot P(w \mid x_1, z) \cdot \Big[P(z \mid x_0) - P(z \mid x_1)\Big] \tag{C9}
\end{equation}

\vspace{3em}

\section*{Appendix D. Conditional average treatment effects (CATE) of ADHD on STEM outcomes by subgroups
}
\addcontentsline{toc}{section}{Appendix D}

\begin{table}[htbp]
\centering
\footnotesize
\renewcommand{\arraystretch}{1.1}
\begin{tabular}{l@{\hskip 8pt}c@{\hskip 6pt}c@{\hskip 6pt}c@{\hskip 10pt}c@{\hskip 6pt}c@{\hskip 6pt}c}
\toprule
 & \multicolumn{3}{c}{\textbf{Science Identity}} & \multicolumn{3}{c}{\textbf{STEM GPA}} \\
\cmidrule(lr){2-4} \cmidrule(lr){5-7}
\textbf{Variables} & \textbf{Mean CATE} & \textbf{SD} & \textbf{n} & \textbf{Mean CATE} & \textbf{SD} & \textbf{n} \\
\midrule
\textbf{Sex} & & & & & & \\
\quad Male & $-$0.038 & 0.087 & 6,840 & $-$0.504 & 0.133 & 7,320 \\
\quad Female & $-$0.034 & 0.084 & 6,910 & $-$0.512 & 0.128 & 7,280 \\[3pt]
\textbf{Race/Ethnicity} & & & & & & \\
\quad White & $-$0.029 & 0.082 & 8,010 & $-$0.521 & 0.134 & 8,410 \\
\quad AI/AN/Pacific Isl. & $-$0.020 & 0.087 & 150 & $-$0.491 & 0.135 & 170 \\
\quad Asian & $-$0.053 & 0.076 & 1,040 & $-$0.528 & 0.11 & 1,090 \\
\quad Black & $-$0.035 & 0.079 & 1,270 & $-$0.431 & 0.097 & 1,370 \\
\quad Hispanic/Latino & $-$0.014 & 0.083 & 2,100 & $-$0.442 & 0.114 & 2,280 \\
\quad Multiracial & $-$0.112 & 0.084 & 1,190 & $-$0.609 & 0.072 & 1,270 \\[3pt]
\textbf{Family SES Quintile} & & & & & & \\
\quad Q1 (Lowest) & $-$0.042 & 0.096 & 2,140 & $-$0.405 & 0.131 & 2,450 \\
\quad Q2 & $-$0.015 & 0.082 & 2,240 & $-$0.502 & 0.126 & 2,370 \\
\quad Q3 & $-$0.040 & 0.075 & 2,360 & $-$0.530 & 0.142 & 2,510 \\
\quad Q4 & $-$0.051 & 0.096 & 2,660 & $-$0.532 & 0.133 & 2,780 \\
\quad Q5 (Highest) & $-$0.032 & 0.078 & 4,350 & $-$0.540 & 0.089 & 4,480 \\[3pt]
\textbf{Highest Educ.\ Expectation} & & & & & & \\
\quad Don't know & $-$0.040 & 0.085 & 2,760 & $-$0.444 & 0.125 & 2,970 \\
\quad High School & 0.015 & 0.087 & 1,380 & $-$0.333 & 0.092 & 1,550 \\
\quad Associate's & 0.007 & 0.072 & 790 & $-$0.474 & 0.085 & 840 \\
\quad Bachelor's & $-$0.012 & 0.061 & 2,440 & $-$0.553 & 0.117 & 2,570 \\
\quad Master's & $-$0.050 & 0.084 & 3,130 & $-$0.560 & 0.099 & 3,270 \\
\quad PhD/Professional & $-$0.069 & 0.086 & 3,250 & $-$0.567 & 0.093 & 3,390 \\
\bottomrule
\end{tabular}
\end{table}


\newpage

\begin{thebibliography}{99}

\bibitem[Athey et~al.(2019)]{Athey2019}
Athey, S., Tibshirani, J., \& Wager, S. (2019).
Generalized random forests.
\emph{The Annals of Statistics}, 47(2), 1148--1178. https://doi.org/10.1214/18-AOS1709

\bibitem[Cinelli \& Hazlett(2020)]{Cinelli2020}
Cinelli, C., \& Hazlett, C. (2020).
Making sense of sensitivity: Extending omitted variable bias.
\emph{Journal of the Royal Statistical Society Series B: Statistical Methodology}, 82(1), 39--67. https://doi.org/10.1111/rssb.12348

\bibitem[Matson et~al.(2010)]{Matson2010}
Matson, J.~L., Mahan, S., Hess, J.~A., \& Fodstad, J.~C. (2010).
Effect of developmental quotient on symptoms of inattention and impulsivity among toddlers with autism spectrum disorders.
\emph{Research in Developmental Disabilities}, 31(2), 464--469. https://doi.org/10.1016/j.ridd.2009.10.014

\bibitem[Meyer(2017)]{Meyer2017}
Meyer, D. (2017).
Bill Gates thinks tech could make inequality worse. But he has faith in robots.
\emph{Fortune}. Retrieved from https://fortune.com/2017/11/15/bill-gates-technology-inequality-robots/

\bibitem[{Office of Special Education Programs}(2015)]{OSEP2015}
Office of Special Education Programs. (2015).
37th annual report to Congress on the implementation of the Individuals with Disabilities Education Act, 2015.
Washington, DC: U.S.\ Department of Education.

\bibitem[Owens(2020)]{Owens2020}
Owens, J. (2020).
Relationships between an ADHD diagnosis and future school behaviors among children with mild behavioral problems.
\emph{Sociology of Education}, 93(3), 191--214. https://doi.org/10.1177/0038040720909296

\bibitem[Pearl(2009)]{Pearl2009}
Pearl, J. (2009).
\emph{Causality: Models, Reasoning, and Inference}.
Cambridge University Press.

\bibitem[Pearl et~al.(2021)]{Pearl2021}
Pearl, J., Glymour, M., \& Jewell, N.~P. (2021).
\emph{Causal Inference in Statistics: A Primer} (Reprinted with revisions).
Wiley.

\bibitem[Plecko \& Bareinboim(2024)]{Plecko2024}
Plecko, D., \& Bareinboim, E. (2024).
Causal fairness analysis: A causal toolkit for fair machine learning.
\emph{Foundations and Trends in Machine Learning}, 17(3), 304--589. https://doi.org/10.1561/2200000106

\bibitem[Plecko et~al.(2025)]{Plecko2025}
Plecko, D., Secombe, P., Clarke, A., Fiske, A., Toby, S., Duff, D., Pilcher, D., Celi, L.~A., Bellomo, R., \& Bareinboim, E. (2025).
An algorithmic approach for causal health equity: A look at race differentials in intensive care unit (ICU) outcomes.
\emph{arXiv preprint arXiv:2501.05197}. https://doi.org/10.48550/arXiv.2501.05197

\bibitem[Saatcioglu \& Skrtic(2019)]{Saatcioglu2019}
Saatcioglu, A., \& Skrtic, T.~M. (2019).
Categorization by organizations: Manipulation of disability categories in a racially desegregated school district.
\emph{American Journal of Sociology}, 125(1), 184--260. https://doi.org/10.1086/703957

\bibitem[Shifrer et~al.(2013)]{Shifrer2013}
Shifrer, D., Callahan, R.~M., \& Muller, C. (2013).
Equity or marginalization? The high school course-taking of students labeled with a learning disability.
\emph{American Educational Research Journal}, 50(4), 656--682. https://doi.org/10.3102/0002831213479439

\bibitem[Shifrer \& Mackin~Freeman(2021)]{Shifrer2021}
Shifrer, D., \& Mackin~Freeman, D. (2021).
Problematizing perceptions of STEM potential: Differences by cognitive disability status in high school and postsecondary educational outcomes.
\emph{Socius: Sociological Research for a Dynamic World}, 7, 2378023121998116. https://doi.org/10.1177/2378023121998116

\bibitem[Wager \& Athey(2018)]{Wager2018}
Wager, S., \& Athey, S. (2018).
Estimation and inference of heterogeneous treatment effects using random forests.
\emph{Journal of the American Statistical Association}, 113(523), 1228--1242. https://doi.org/10.1080/01621459.2017.1319839

\bibitem[Zhang \& Bareinboim(2018)]{Zhang2018}
Zhang, J., \& Bareinboim, E. (2018).
Fairness in decision-making---the causal explanation formula.
\emph{Proceedings of the AAAI Conference on Artificial Intelligence}, 32(1). https://doi.org/10.1609/aaai.v32i1.11564

\bibitem[Zhang et~al.(2019)]{ZZhang2019}
Zhang, Z., Kim, H.~J., Lonjon, G., \& Zhu, Y. (2019).
Balance diagnostics after propensity score matching.
\emph{Annals of Translational Medicine}, 7(1), 16. https://doi.org/10.21037/atm.2018.12.10

\end{thebibliography}
\end{document}